\input{aipcheck}
\pdfoutput=1

\documentclass[
    ,final            
  ,numberedheadings 
  ]
  {aipproc}

\layoutstyle{8x11single}

\usepackage{slashed}

\def\be{\begin{equation}}
\def\ee{\end{equation}}
\def\bea{\begin{eqnarray}}
\def\eea{\end{eqnarray}}
\def\ls{\mathrel{\lower4pt\vbox{\lineskip=0pt\baselineskip=0pt
           \hbox{$<$}\hbox{$\sim$}}}}
\def\gs{\mathrel{\lower4pt\vbox{\lineskip=0pt\baselineskip=0pt
           \hbox{$>$}\hbox{$\sim$}}}}

\newcommand{\gsim}{\lower.7ex\hbox{$\;\stackrel{\textstyle>}{\sim}\;$}}
\newcommand{\lsim}{\lower.7ex\hbox{$\;\stackrel{\textstyle<}{\sim}\;$}}

\newcommand{\neu}[1]{\ensuremath{\tilde{\chi}_{#1}^0}}

\begin{document}

\title{Higgsino Dark Matter and the Cosmological Gravitino Problem}


\author{Kuver Sinha}{
  address={Mitchell Institute for Fundamental Physics and Astronomy, 
  Texas A \& M University, College Station, TX}
}


\begin{abstract}

We motivate Higgsino dark matter from a solution to the cosmological moduli/gravitino problem. Cosmological moduli/gravitino should be heavy enough to decay before the onset of Big Bang Nucleosynthesis, and this requirement typically forces gauginos to have masses above a TeV in Type IIB compactifications. Higgsinos emerge as the viable sub-TeV dark matter candidates if anomaly and modulus mediated contributions to supersymmetry breaking are both competitive. 
Obtaining the correct relic density in this mass range forces Higgsinos to be produced non-thermally from the decay of a modulus. We outline constraints arising from indirect and direct detection experiments in this context, as well as theoretical constraints such as the overproduction of dark matter from gravitino decay.

\end{abstract}

\maketitle


\section{Introduction}

In the $R$- parity conserving Minimal Supersymmetric Standard Model (MSSM), the lightest supersymmetric particle (LSP) is the dark matter candidate. A neutralino LSP that is purely Higgsino has been motivated recently from a cosmological perspective \cite{Allahverdi:2012wb}, in addition to the usual motivation from naturalness considerations \cite{Hall:2011aa, Baer:2012uy, Papucci:2011wy} \footnote{
In Natural Supersymmetry, Higgsinos are the LSP since all superpartners are taken to be heavy unless they play a role in the naturalness of the scale of Electro Weak Symmetry Breaking (EWSB). From the perspective of naturalness, the appearance of $\mu$ in the tree-level relation for $m_Z$ implies that $\mu$ cannot be too large. Since the Higgsino masses are directly proportional to $\mu$, they cannot be too heavy.}. The cosmological perspective taken has been as follows: in simple calculable solutions to the cosmological moduli/gravitino problem \cite{BBN}, in the setting of mixed modulus/anomaly mediated supersymmetric models in type IIB string theory, the Higgsinos emerge as the only viable dark matter candidates with sub-TeV masses. It is this cosmological motivation, relating to the moduli problem, that will be discussed in detail in the present article \footnote{We note that Higgsino dark matter has cosmological motivations other than the ones considered here, for example from baryogenesis \cite{Blum:2012nf}.}. 

A Higgs boson with a mass of around $125$ GeV \footnote{Recent experiments at the LHC have provided strong hints of a Higgs-like particle at $\sim 125$ GeV \cite{LHCHiggs}.} supports (but does not in itself motivate) this approach to Higgsino dark matter. A $125$ GeV Higgs is obtained, as we will show, in an explicit supersymmetric model with mixed modulus/anomaly mediation, where $(a)$ the gravitino is heavy enough to decay before Big Bang Nucleosynthesis (BBN) and $(b)$ Higgsinos are the dark matter and have correct relic density. 

Before we elaborate on this cosmological approach further, we briefly survey some proposals for obtaining the correct relic density with Higgsino dark matter. If the lightest neutralino is pre-dominantly Higgsino, with mass in the sub-TeV region but larger than $m_W$, the annihilation rate is larger than the nominal value $\langle \sigma_{\rm ann} v \rangle = 3 \times 10^{-26}$ ${\rm cm^3 s^{-1}}$, thus resulting in an insufficient thermal relic abundance. Light Higgsinos with mass less than $m_W$, suppressing $\neu{1} \neu{1} \rightarrow W^+ W^-$, have been studied by several authors with a view to obtaining the correct relic density \cite{Drees:1996pk}, and satisfying constraints from direct searches \cite{Hisano:2004pv}. For Higgsinos larger than $\sim 1$ TeV, usual thermal freezeout can give the correct relic density, but this is not motivated either from a naturalness viewpoint, or from observational accessibility. 

We will be interested in Higgsinos in the intermediate mass range, especially with mass $\mathcal{O}(100-300)$ GeV. The most natural option to obtain the correct relic density is to rely on non-thermal production of Higgsino dark matter \cite{Moroi:1999zb}, which also fits in perfectly with the cosmological motivation to be outlined \footnote{Another option for obtaining the correct relic density in this range is to assume other sources for dark matter in addition to Higgsinos, as described elsewhere in these proceedings \cite{{Baer:2012uy}}.}.  

\subsection{Cosmological Moduli/Gravitino and Higgsino Dark Matter}

The cosmological moduli problem has been discussed quite extensively in the literature, from the point of view of hidden sector supersymmetric models with gravity mediation, as well as in the context of string moduli stabilization. Briefly, the problem is as follows. 
In the early universe moduli are typically displaced form the minimum of their potential, oscillate, and behave like non-relativistic matter once the Hubble expansion rate drops below their mass. The moduli are long lived because their couplings to other fields are gravitational, and their late decay can spoil the successful predictions of BBN.

Even when moduli decay sufficiently early not to affect BBN, several issues arise - their decay dilutes previously existing dark matter and baryon asymmetry. However, this also has the potential to give rise to very interesting physics, for example models of post-sphaleron baryogenesis \cite{Allahverdi:2010im} and non-thermal sneutrino dark matter \cite{Dutta:2009uf}. Moreover, modulus decay gives a natural explanation of the dark matter-barogenesis coincidence problem in the context of Cladogenesis \cite{Allahverdi:2010rh}, which is an alternative to asymmetric dark matter scenarios.

The case when the final decaying particle is a gravitino is even more interesting, since it has direct consequences for the supersymmetric spectrum. Gravitinos heavier than ${\cal O}(40)$ TeV have a lifetime shorter than $0.1$ s and decay before the onset of BBN. This results in a considerable relaxation as the gravitino abundance will not be subject to tight BBN bounds, thus evading the cosmological gravitino problem. In effective supergravity, the masses of the Bino and Wino are sensitive to the mass of the gravitino $m_{3/2}$  \cite{Kaplunovsky:1993rd}, and thus solving the cosmological gravitino problem has direct consequences for low-energy particle physics, especially the identity and mass of dark matter. 

The exact consequences depend on the overall suppression of the lightest neutralinos with respect to the gravitino, and the relative suppressions of the neutralinos compared to each other. For example, both in the case of pure anomaly mediation as well as modulus mediation in the case of the $G_2$-MSSM 
models \cite{Acharya:2008zi} the LSP is a Wino in the mass range of $\mathcal{O}(100-300)$ GeV for a gravitino with mass ${\cal O}(40)$ TeV.

Type IIB modulus mediation models offer another alternative, due to the fact that one can control the relative preponderance of modulus and anomaly mediation contributions. One typically has Bino and Wino masses above a TeV for $m_{3/2} > 40$ TeV, but the Higgsino mass depends on the $\mu$ parameter, which can be reduced as anomaly mediated contributions to supersymmetry breaking become relatively more important. As a result, if we demand that the dark matter particle has a mass in the sub-TeV region, the Higgsino becomes a more natural dark matter candidate in these models.

Moreover, the decay of the Kahler modulus responsible for supersymmetry breaking itself provides a non-thermal setting in which the Higgsino dark matter can have mass in the interesting range, as well as satisfy the relic density. These scenarios represent, therefore, calculable examples with the minimal ingredients to solve the moduli problem, obtain Higgsino dark matter with correct relic density and, as we shall see, a Higgs mass that is also in the correct ballpark.

There is another angle to the study undertaken here. Regardless of the exact UV theory under which Higgsinos arise as dark matter, non-thermal production is the only way for them to satisfy the relic density in the intermediate mass range, and it is important to understand the constraints on such production. The constraints discussed are both theoretical as well as experimental. We find that for annihilation rate to be compatible with bounds from the Fermi Gamma-Ray Telescope~\cite{fermi}, the modulus decay should reheat the universe to a temperature $T_{\rm d} \sim {\cal O}({\rm GeV})$.
On the other hand, a theoretical requirement is that the branching ratio for modulus decay to the gravitino is $\ls {\cal O}(10^{-5})$, so that the decay of gravitinos thus produced does not lead to dark matter overproduction. In fact, this theoretical requirement will require us to go beyond the explicit example considered here.

The paper is organized as follows. In Section \ref{mainsection}, we describe Higgsino dark matter in a non-thermal setting using an explicit example. In Section \ref{gravitino}, we describe the constraints on gravitino decay in more general settings to prevent overproduction of dark matter. We end with our conclusions.

\section{Higgsino Dark Matter and Non-thermal Physics: An Explicit Example} \label{mainsection}


In the following subsections, we first show how Higgsinos emerge as the viable LSP with sub-TeV mass in type IIB scenarios, then discuss the non-thermal history coming from the decay of a K\"ahler modulus, and finally present our results. We then describe the role of the gravitinos and issues with dark matter produced by their decay.

The standard scenario of KKLT compactification \cite{Kachru:2003aw}, with the Kahler modulus reheating the universe around $\mathcal{O}(1)$ GeV, offers an example of the kind of non-thermal scenario that we want to describe. The emergence of Higgsinos as dark matter candidates hinges on the relative calibration between anomaly mediation and modulus contributions to GUT scale gaugino masses. This class of models offer the scope of such calibration. Moreover, the modulus sector is well-defined. 

We will come to the conclusion that although this example displays all the desirable features for Higgsino dark matter, it does not work fully due to the overproduction of dark matter from gravitino decay. We take up this issue in more generality in the next section.

\subsection{Obtaining Higgsino LSP}

We first outline the general ingredients necessary for obtaining Higgsino LSP in the context of the cosmological moduli/gravitino problem, and then work out our concrete example.

Obtaining a TeV scale spectrum in the observable sector, while keeping gravitinos heavy enough to avoid BBN bounds, requires a hierarchy between the gravitino and the other superpartner masses. For gauginos, at least, this is obtained quite generally in Type IIB compactifications, since their masses obey
\be \label{hierarchy}
M_{\tilde g} \, \approx \, \frac{m_{3/2}}{\ln{(M_{\rm P}/m_{3/2})}} \,\,.
\ee
We note that this hierarchy has been argued to be generic \cite{Conlon:2006us}, and depends only on general ingredients \footnote{Specifically, $(a)$ a K\"ahler modulus $T_a$ stabilized by non-perturbative effects $(b)$ complex structure moduli stabilized by fluxes $(c)$ visible sector on $D7$ branes.}, not on the specifics of the stabilization scheme.

Parametrizing the relative contributions of anomaly and modulus mediation by $\alpha \equiv m_{3/2}/M_0\ln(M_{\rm P}/m_{3/2})$, where $M_0$ is the modulus mediated contribution at the GUT scale, one obtains the ratios of the gaugino masses as follows \cite{Choi:2006im, Choi:2005uz}
\bea \label{alphagauginos}
&M_3 &: M_2 : M_1  \nonumber \\
&\sim & (1 - 0.3 \alpha)g_3^2 : (1+ 0.1 \alpha)g_2^2 : (1+ 0.66 \alpha)g_1^2 \,\, .
\eea
In the above, $g_{1,2,3}$ are the gauge coupling constants \footnote{The mass hierarchy between the scalar masses and the gravitino mass is more model dependent, and depends on the curvature properties of the underlying K\"ahler manifold.}.

For $\alpha \rightarrow 0$ the anomaly mediated contribution vanishes and the Bino is the LSP and has mass typically above $\mathcal{O}(1)$ TeV for $m_{3/2} > 40$ TeV, as is clear from Eq.~\ref{alphagauginos}. For the Higgsinos, there is additional freedom. Increasing $\alpha$ lowers the gluino mass, while the Bino and the Wino become heavier. Lowering the gluino mass in turn lowers the low-scale value of $m^2_{H_u}$ due to the top Yukawa coupling. Since the value of the Higgsino mass parameter $\mu$ depends on the low-scale value of $m^2_{H_u}$, which is mainly driven by the gluino mass \cite{Endo:2005uy}, Higgsinos become more preferred as the LSP as anomaly contributions become stronger. 

In fact, demanding that the mass of the dark matter candidate has to be less than $\mathcal{O}(1)$ TeV, Higgsinos become the only viable candidates.


The above considerations are quite general. The case of mirage mediation in the context of KKLT compactification serves as a specific example. 

\begin{figure}[tb]
\includegraphics*[width=0.60\columnwidth]{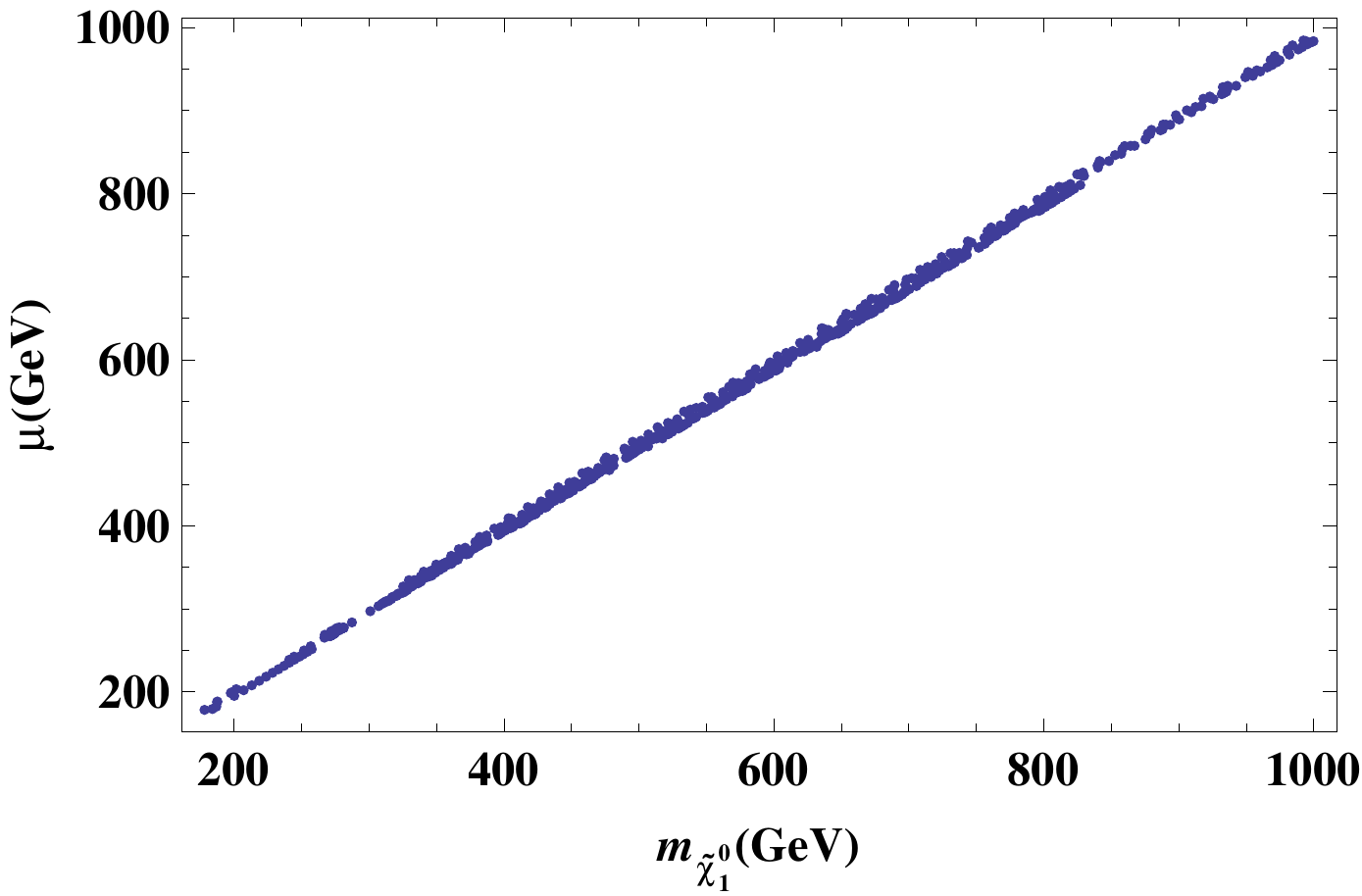}
\vspace*{-.1in}
\caption{$\mu$ versus LSP mass. For sub-TeV LSP, the dark matter is always a Higgsino.}
\label{massneutralinomu}
\end{figure}

The superpotential of the modulus sector consists of a flux term that fixes complex structure moduli, and a non-perturbative piece that fixes the K\"ahler modulus. The K\"ahler potential is given by
\be
K = -3\ln{(T+\overline{T})} + (T+\overline{T})^{-n_m}\Phi \Phi^{\dagger}, 
\ee
where $\Phi$ denotes matter fields and $n_m$ are the modular weights. In the above, $T$ denotes the K\"ahler modulus. The input parameters fixing the GUT scale masses are $m_{3/2}$, $\alpha$, $n_m$, and ${\rm tan}\beta$. For our case study, we choose $n_m = 1/2$ for all matter fields and ${\rm tan}\beta = 50$. The general conclusions hold for other values of $n_m$ and ${\rm tan} \beta$. Scanning over $0.1 < \alpha < 1.6$, and $m_{3/2} > 40$ TeV, one finds that for LSP mass below $\sim 1$ TeV, the LSP is always a Higgsino. We plot $\mu$ against LSP mass in Figure \ref{massneutralinomu}. Clearly, the linear relaion between $\mu$ and the LSP mass shows a Higgsino dark matter candidate. 

The scalar spectrum has the suppression given in Eq.~(\ref{hierarchy}) with respect to the gravitino mass. It is instructive to note that when the stops are themselves hierarchically related to the gravitino as in this case, $m_h \sim 125$ GeV is compatible with heavy gravitinos that decay before the onset of BBN. Since the one-loop correction to the Higgs mass depends logarithmically on $m_{3/2}$, a heavier Higgs is preferred by the cosmologically safe region.

\begin{figure}[tb]
\includegraphics*[width=0.60\columnwidth]{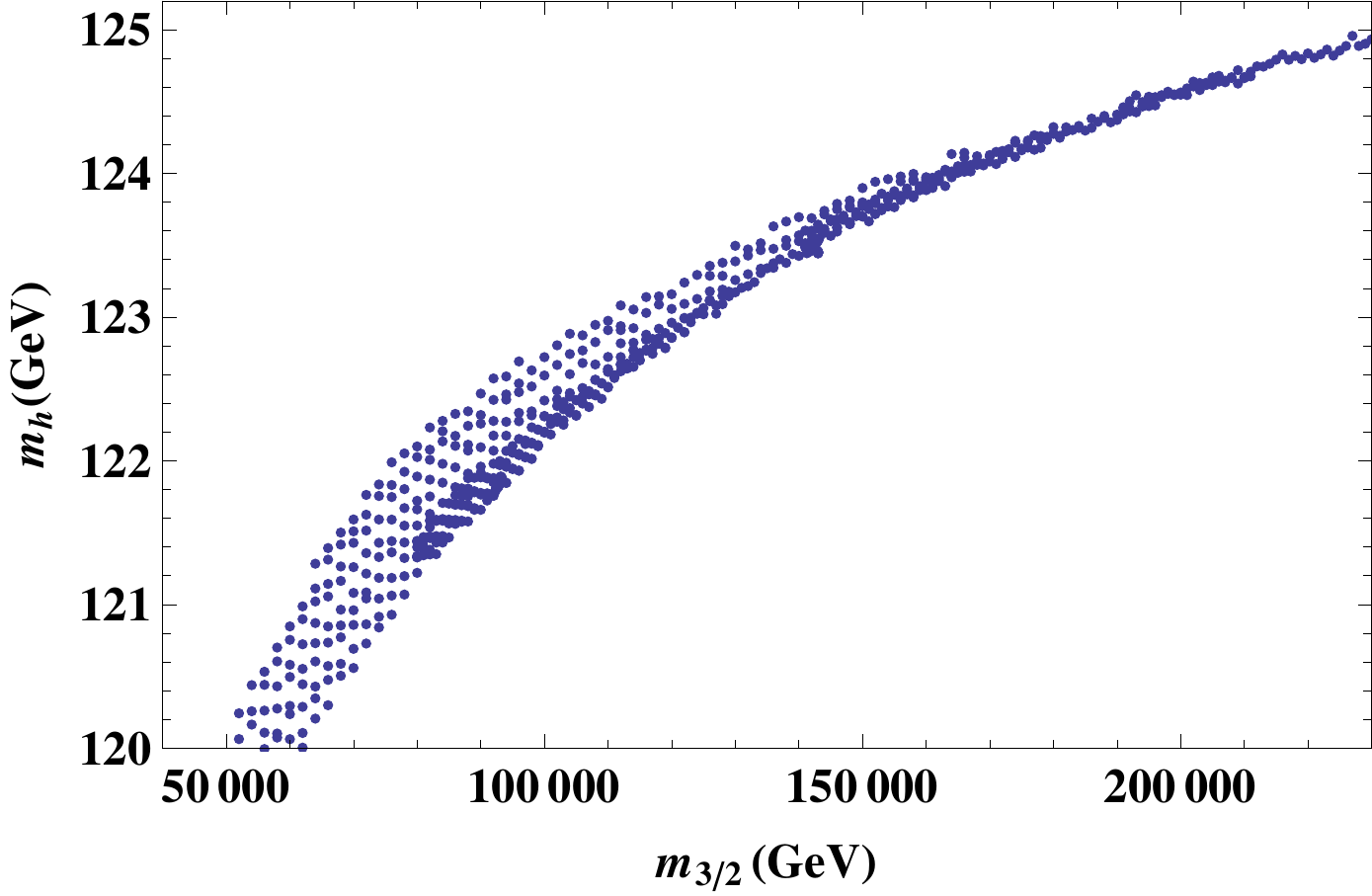}
\vspace*{-.1in}
\caption{Higgs mass versus gravitino mass. We choose $n_m =1/2$ and ${\rm tan}\beta = 50$, and scan over $\alpha$ and $m_{3/2}$. Heavy gravitinos decaying before the onset of BBN are typically compatible with a Higgs mass above 120 GeV. A similar behavior is obtained for other values of ${\rm tan}\beta$ and $n_m$}
\label{higgsmassgravitinomass}
\end{figure}

We plot the dependence of the Higgs mass on the gravitino mass in Figure \ref{higgsmassgravitinomass}.



\subsection{Non-thermal History, Relic Density, and Constraints from Indirect Detection}


Having understood the origin of Higgsino LSPs, we go on to describe non-thermal production of dark matter and the experimental constraints on these scenarios.

If a modulus $\phi$, which for our purposes will be the real part of the K\"ahler modulus $T$, decays and reheats the universe at a scale below the dark matter freeze-out temperature $T_{\rm f} \sim m_\chi/25$, then one must consider non-thermal production of dark matter and the ensuing non-thermal history. The decay width of $\phi$, which has couplings of gravitational strength to the visible sector fields, is
\be \label{decaywidth}
\Gamma_{\phi} = \frac{c}{2\pi} \frac{m_\phi^3}{M^2_{\rm P}} ,
\ee
where $c$ depends on the couplings of the decaying field. For moduli fields, we typically have $c \sim 0.1-1$. 

The modulus decays when $H \simeq \Gamma_{\phi}$, with $H$ being the Hubble expansion rate of the universe. Modulus decay reheats the universe to the following temperature
\begin{eqnarray} \label{Td}
T_{\rm d} \simeq
(5 ~ {\rm MeV}) ~ c^{1/2} \left(\frac{10.75}{g_*}\right)^{1/4} \left(\frac{m_\phi}{100~{\rm TeV}}\right)^{3/2} \, , \nonumber \\
\end{eqnarray}
where $g_*$ is the total number of relativistic degrees of freedom at $T_{\rm d}$ ($g_* = 10.75$ for $T_{\rm d} \sim {\cal O}({\rm MeV})$).

One can also treat the case where $\phi$ is the gravitino similarly. Then, $c$ can be computed explicitly since supersymmetry fixes the couplings of the gravitino to the visible sector. One has a maximal value of $c \sim 1.5$ in this case \cite{Moroi:1995fs}. Gravitinos that have a lifetime shorter than $0.1$ s decay before the onset of BBN and avoid any conflict with its successful predictions. Such a lifetime corresponds to $T_{\rm d} \gs 3$ MeV, which requires that $m_{3/2} \gs {\cal O}(40)$ TeV from Eq.~(\ref{Td}).

The dark matter relic density from modulus decay is given by
\be \label{dmdensity}
{n_\chi \over s} \approx 5 \times 10^{-10} ~ \left({1 ~ {\rm GeV} \over m_\chi} \right) ~ {3 \times 10^{-26} ~ {\rm cm^3 ~ s^{-1}} \over \langle \sigma_{\rm ann} v \rangle_{\rm f}} ~ \left({T_{\rm f} \over T_{\rm d}}\right) ,
\ee
where
$\langle \sigma_{\rm ann} v \rangle_{\rm f}$ is the annihilation rate at the time of freeze-out.

We are interested in the case of Higgsino dark matter, which mainly annihilates into heavy Higgs bosons, $W$ bosons and $t$ quarks via $S$-wave annihilation if $m_\chi$ has necessary phase space for these particles to be produced. Since the annihilation occurs through $S$-wave, the annihilation rate at the freeze-out time is essentially the same as that at the present time. The latter is in turn constrained by the gamma ray flux from dwarf spheriodal galaxies~\cite{fermi}:
\begin{eqnarray} \label{fermi}
&& \langle \sigma_{\rm ann} v \rangle_f \ls 10^{-25} ~ {\rm cm}^3 ~ {\rm s}^{-1} ~ ~ ~ ~ ~ ~ ~ ~ ~ ~ m_\chi = 100 ~ {\rm GeV} \, , \nonumber \\
&& \langle \sigma_{\rm ann} v \rangle_f \ls 3 \times 10^{-24} ~ {\rm cm}^3 ~ {\rm s}^{-1} ~ ~ ~ ~ ~ m_\chi = 1 ~ {\rm TeV} .
\end{eqnarray}

According to \cite{Hooper:2012sr}, the constraint on the annihilation cross-section from the gamma-ray flux from the galactic center region  is similar for the above neutralino masses to a core of $1$ kpc in the $b \overline{b}$ final states. The constraint on the annihilation cross-section becomes about $4 \times 10^{-26}~ {\rm cm}^3 ~ {\rm s}^{-1}$ for $m_{\tilde\chi} = 100$ GeV for the NFW profile without any core.

A note on other potential constraints: $(a)$ Explaining the anomaly observed by PAMELA requires much larger cross-section \cite{Cirelli:2008pk}, but the explanation of that anomaly can be pulsars \cite{Yuksel:2008rf}. $(b)$ The bounds on the cross section from dark matter annihilation to neutrinos at the galactic center, obtained by IceCube, are weaker by few orders of magnitude \cite{IceCube:2011ae}.

Obtaining the correct DM abundance, see Eq.~(\ref{dmdensity}), translates into a range for the reheat temperature:
\begin{eqnarray} \label{td2}
T_{\rm d} \gs 0.4-1.6 ~ {\rm GeV} ~ ~ ~ ~ ~ ~ m_\chi = 100 ~ {\rm GeV}-1 ~ {\rm TeV} \, .
\end{eqnarray}

This in turn translates into bounds for the modulus mass. For $T_{\rm d} \sim {\cal O}({\rm GeV})$, the corresponding modulus mass is found from Eq.~(\ref{Td}) to be
\be\label{mphi}
m_\phi \sim \mathcal{O}(1000) ~ {\rm TeV}\,,
\ee
with the exact value depending on the decay modes of the modulus.

In the particular model under consideration, the mass of the K\"ahler modulus is in fact in the requisite range.

\subsection{Results}

In the previous two subsections, we have first explored how Higgsino LSPs arise in the context of mixed modulus/anomaly mediation, and then studied constraints for non-thermal production of Higgsinos coming from the relic density and Fermi Gamma-Ray Telescope bounds. In this subsection, we study the specific case of mirage mediation in KKLT.

In Table \ref{KKLTbenchmark}, a few benchmark points of non-thermal Higgsino dark matter in the mirage mediation model are shown. The table also includes masses of the gluino and stops, and the Higgs mass $m_h$ for these points.

\begin{table}[h]
\caption{Some benchmark points of non-thermal Higgsino dark matter for mirage mediation model in the context of KKLT compactification. The input parameters are $\alpha$ and $m_{3/2}$. The modular weights are fixed to be $n_m = 1/2$, and ${\rm tan}\beta = 50$.
All masses are in GeV.}
\label{KKLTbenchmark}
\begin{tabular}{c c | c c c c c c c c c c} \hline \hline


$\alpha$  & $m_{3/2}$  & $m_h$  & $m_{\tilde{\chi}^0_1}$ & $m_{\tilde{\chi}^0_2}$ & $m_{\tilde{g}}$ & $m_{\tilde{t}_1}$ & $m_{\tilde{t}_2}$  & $\langle \sigma_{\rm ann} v \rangle_{0}$ & $\langle \sigma_{\rm ann} v \rangle_{f}$ & $T_{\rm f}/T_{\rm d}$ &  $\sigma_{\tilde{\chi}^0_1-p}$            \\ \hline \hline
$1.49$ & $143 \cdot 10^3$  & $123.5$  & $248.4$ & $250.8$  & $3828$ & $2441$ & $2781$  & $1.49 \cdot 10^{-25}$ & $1.63 \cdot 10^{-25}$  & $\sim 6$ & $5 \cdot 10^{-10}$ \\

$1.46$  & $200 \cdot 10^3$ & $124.5$ & $258.9$& $260.6$ & $5536$ & $3564$ &$3991$ & $1.38 \cdot 10^{-25}$  & $1.52 \cdot 10^{-25}$ & $\sim 3.4$ & $1.4 \cdot 10^{-10}$  \\

$1.44$ & $232 \cdot 10^3$ & $125$ & $306$ & $308$  & $6505$ & $4197$  &$4677$ & $1.01 \cdot 10^{-25}$ & $1.01 \cdot 10^{-25}$ & $\sim 3.2$ & $8.9 \cdot 10^{-11}$  \\   \hline \hline

\end{tabular}
\end{table}

Firstly, the annihilation rates at present $\langle \sigma_{\rm ann} v \rangle_0$ and at the time of freeze-out $\langle \sigma_{\rm ann} v \rangle_f$ are given to check compatibility with bounds coming from Fermi. We see that for all of the points shown, $\langle \sigma_{\rm ann} v \rangle_0$ satisfies the Fermi bounds. Of the three points presented in Table \ref{KKLTbenchmark}, the first two cases satisfy the constraint from the dwarf spheroidals and the flux arising from galactic center region with a core of $1$ kpc. The third satisfies the constraints from dwarf spheroidals, flux from the galactic center with and without any core for NFW profile.

Secondly, we check that the dark matter relic density is satisfed  by the non-thermal production. Dark matter annihilation at the freeze-out occurs mostly through $S$-channel and a coannhilation component. The latter arises due to the fact that masses of the second lightest neutralino and chargino are close to LSP mass $m_{{\tilde \chi}^1_0}$.
The dark matter content of the universe is obtained by multiplying the $\langle \sigma_{\rm ann} v \rangle_f$ by $T_{\rm f}/T_{\rm d}$.
We have taken $c=0.4$ to calculate $T_{\rm d}$, which is the leading order value appearing in the decay width of the modulus in this particular example, in the limit of $\alpha = 1$ corresponding to zero dilaton-modulus mixing in the gauge kinetic function. But $c$ can also be $\sim 1$ depending on relative contributions of the modulus and dilaton in the gauge kinetic function, and $T_{\rm d}$ can be $\mathcal{O}(1-2)$ of its central value. With this taken into account, it is seen that $T_{\rm f}/T_{\rm d}$ is in the right range to yield the correct dark matter content of the universe.

\begin{figure}[tb]
\includegraphics*[width=0.60\columnwidth]{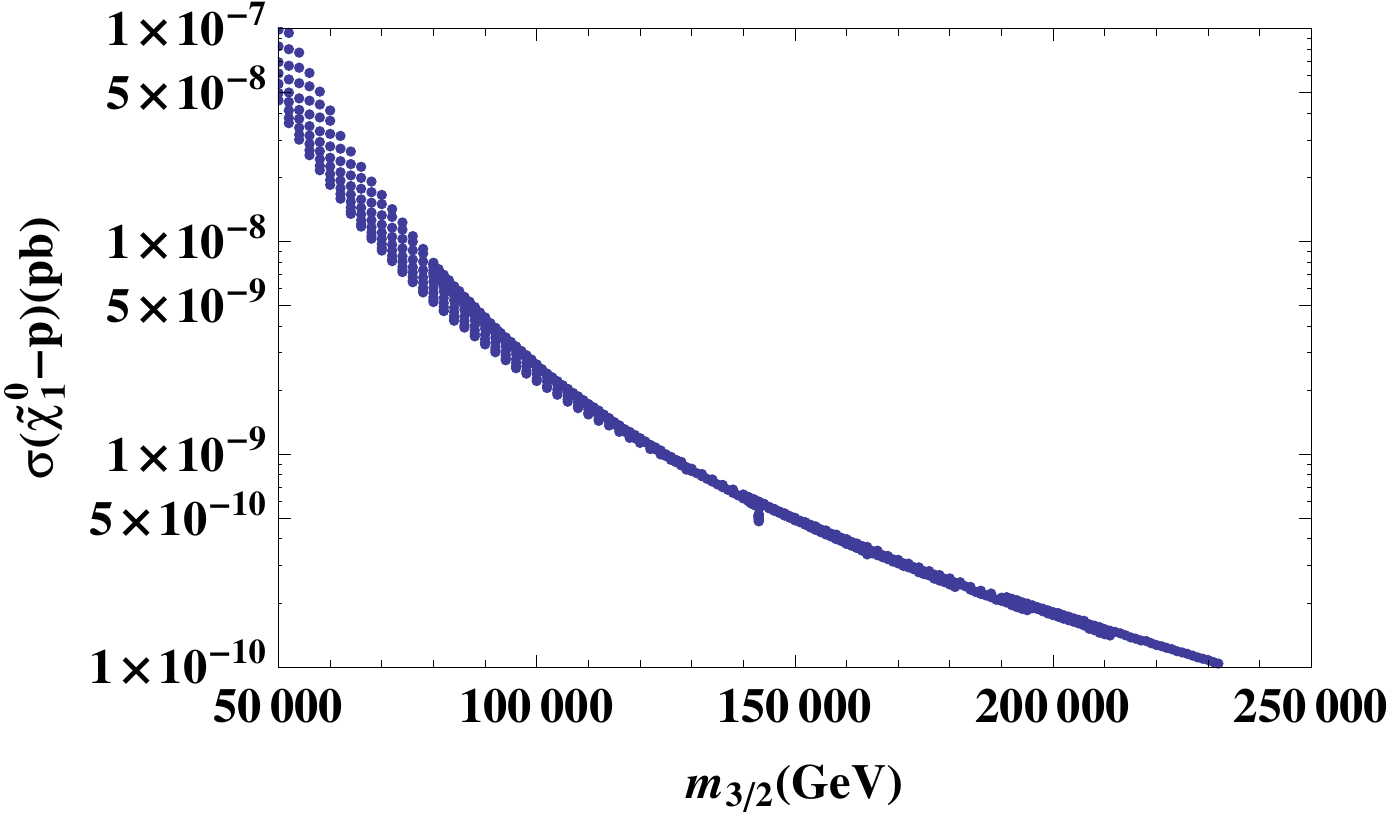}
\vspace*{-.1in}
\caption{Spin independent scattering cross section versus gravitino mass. For values of $m_{3/2}$ that are compatible with 125 GeV Higgs, see Fig. 1, $\sigma_{{\tilde \chi}^1_0-p}$ satisfies the experimental data.}
\label{protonneutralinocrosssectiongravitino}
\end{figure}

Thirdly, we also show the value of $\sigma_{\tilde\chi^0_1-p}$ for these points, which are well allowed by the experimental data \cite{direct}. The spin independent scattering cross section $\sigma_{{\tilde \chi}^0_1-p}$ is plotted for various values of the gravitino mass in Figure \ref{protonneutralinocrosssectiongravitino}. Since larger gravitino mass is correlated to a larger heavy Higgs ($H$) mass in this model, $\sigma_{\tilde \chi^0_1-p}$ becomes smaller as $m_{3/2}$ increases. This is compatible with cosmologically safe region and $m_h \sim 125$ GeV. The current bound on the cross section is $\sim 2 \times 10^{-9}$ pb for a dark matter mass of $55$ GeV~\cite{direct}, and relaxes as dark matter mass increases.

It is interesting to calculate the level of fine-tuning for the $m_h = 125$ GeV point in the table. A robust estimator of fine-tuning may be obtained from \cite{Barbieri:1987fn, Anderson:1994dz}. The UV parameters of our model are $\alpha$ and $m_{3/2}$, and the fine-tuning of the Higgs mass and $\mu$ with respect to them are as follows:
\bea
\Delta_{h,\alpha} = \frac{\partial \ln {m_{h}}}{ \partial \ln{\alpha}} = 5.7 \nonumber \\
\Delta_{h, m_{3/2}} = \frac{\partial \ln {m_{h}}}{ \partial \ln{m_{3/2}}} = 2.1 \nonumber \\
\Delta_{\mu, \alpha} = \frac{\partial \ln {\mu}}{ \partial \ln{\alpha}} = 6034 \nonumber \\
\Delta_{\mu, m_{3/2}}  = \frac{\partial \ln {\mu}}{ \partial \ln{m_{3/2}}} = 16 
\eea

\subsection{Role of the Gravitino and Constraint on Production}

In the previous subsections, we have presented results for the example of mirage mediation in type IIB theories, where non-thermal Higgsino LSPs satisfy the relic density and bounds from both direct and indirect detection experiments. The non-thermal history is provided by the decay of the modulus $\phi$, which is the real part of the K\"ahler modulus.

We now turn to a discussion of the role of the gravitino in these scenarios.

Firstly, as mentioned before, the gravitino must decay before BBN; in fact, that sets the scale of the low-energy spectrum.

Secondly, it is important to note that the gravitino is the last particle to decay, not the modulus $\phi$. This will be the case in general: as we have seen, the requisite temperature for modulus decay is around $1$ GeV, while the gravitino decays around $\mathcal{O}(1)$ MeV. This has certain consequences. A source of non-thermal Higgsino production, in addition to that coming from $\phi$ that has been studied already, is now provided by the gravitino. Since the gravitino decays at a temperature $\ll {\cal O}({\rm GeV})$, and dark matter annihilation rate must satisfy the Fermi bounds~(\ref{fermi}), annihilation is very inefficient at this time. Thus, the density of LSPs produced from gravitinos is the same as the density of gravitinos (since $R$-parity is conserved). Therefore, we require
\begin{eqnarray} \label{gravdens}
{n_{3/2} \over s} \ls 5 \times 10^{-10} ~ \left({1 ~ {\rm GeV} \over m_\chi}\right) \, .
\end{eqnarray}

Now, the density of gravitinos is in turn set by their production via thermal and non-thermal processes in the early universe. Modulus decay dilutes gravitinos that were produced in the prior epochs (e.g., during inflationary reheating) by a huge factor. Thermal gravitino production after modulus decay is highly suppressed due to the low decay temperature $T_{\rm d} \sim {\cal O}({\rm GeV})$.

Therefore, the  density of gravitinos is given by what is produced directly from modulus decay $\phi \rightarrow {\tilde G} {\tilde G}$, and it is $(n_{3/2}/s) = {\rm Br}_{3/2} (3 T_{\rm d}/4 m_\phi)$, where ${\rm Br}_{3/2}$ is the branching ratio for $\phi \rightarrow {\tilde G} {\tilde G}$ process. From Eqs.~(\ref{Td}), we then find
\begin{eqnarray} \label{gravbr}
{n_{3/2} \over s} \sim 5 \times 10^{-8} ~ \left(\frac{m_\phi}{100\, {\rm TeV}}\right)^{1/2} ~ {\rm Br}_{3/2} \, .
\end{eqnarray}

For the typical value of $m_\phi$ given in~(\ref{mphi}) and $100~{\rm GeV} \leq m_\chi \leq 1$ TeV, Eqs.~(\ref{gravdens},\ref{gravbr}) yield the following absolute upper bound:
\be \label{brconst}
{\rm Br}_{3/2} \ls 10^{-5} .
\ee
Any successful scenario for non-thermal Higgsino production from modulus decay must satisfy this limit. 

The above bound represents the case where dark matter overproduction from gravitinos is avoided by suppressing the production of gravitinos from the modulus $\phi$ in the first place. We note that this does not obviate the need for the gravitinos that \textit{are} produced to decay before BBN; in particular, the gravitino mass must still be $\mathcal{O}(40)$ TeV.

In the KKLT model discussed above, the partial decay rate for $\phi \rightarrow {\tilde G} {\tilde G}$ is $\Gamma_{3/2} = m^3_\phi/288 \pi M^2_{\rm P}$. Then, after using Eq.~(\ref{decaywidth}), we find that ${\rm Br}_{3/2} \sim {\cal O}(10^{-2})$. This implies that gravitino decay will overproduce Higgsinos by 3 orders of magnitude in this model, see Eq.~(\ref{gravbr}). The main reason for obtaining such a large ${\rm Br}_{3/2}$ is that modulus decay to helicity $\pm 1/2$ gravitinos is not helicity suppressed in the KKLT model \cite{Dine}.

\section{General Conditions for Suppressing Gravitino Production} \label{gravitino}

In the previous section, we have seen that the specific example presented does not completely work. The decay of the gravitinos that are directly produced from modulus decay overproduces dark matter. This is a direct consequence of the couplings between the modulus and the helicity $\pm 1/2$ components of the gravitino, which are in turn set by the underlying K\"ahler geometry of the effective $D=4,~N=1$ supergravity theory. We discuss this in more detail here.

The problem can be overcome if the modulus $\phi$ does not dominate the energy density of the universe when it decays. In such a case, the right-hand side of Eqs.~(\ref{gravbr}) and~(\ref{brconst}) will be multiplied by $f_\phi$ and $f^{-1}_\phi$, respectively, where $f_\phi$ is the ratio of the energy density in $\phi$ to the total energy density of the universe at the time of decay. For $f_\phi < 10^{-3}$, the abundance of gravitinos will be suppressed to safe levels.

Alternatively, one can seek conditions for suppressing gravitino production from modulus decay. Here we briefly outline general conditions for such a suppression, and stress that the main ingredients of a successful scenario for non-thermal Higgsino production presented above should also hold in cases where the gravitino production is suppressed.

The decay of a modulus to other fields depends on the interaction terms in the Lagrangian, and the requirement for suppressing decay to gravitinos will be reduced to a set of constraints in the effective theory. To have a more concrete demonstration of what kinds of constraints may emerge, we choose to work in effective supergravity, with a modulus coupling to the visible sector through the gauge kinetic function. This is the scenario in the class of Type IIB models discussed above.

In general, one can consider a scenario with multiple moduli $\phi_i$, with the decaying modulus appearing in the gauge kinetic function. The normalized eigenstates $\phi_n$ are given by
\be \label{norm}
(\phi)_i \,\, = \,\, \sum_j \,\, C_{ij} \,(\tau_n)_j \,\,\,,
\ee
where the $C_{ij}$ are eigenvectors of the matrix $K^{-1} \,\partial^2 V$. For simplicity, we will assume diagonal $C_{ij}$ with entries $C_i$. The partial widths for modulus decay to gauge fields, gauginos and helicity $\pm 1/2$ gravitinos are
\bea \label{GammaTGravitino}
\Gamma_{\phi_i \rightarrow g g} & = &  \frac{N_g}{128\pi} ~ \frac{1}{\tau^{2}} \, C_i^2 \frac{m_{\phi_i}^3}{M^2_{\rm P}} \nonumber \\
\Gamma_{\phi_i \rightarrow \tilde{g} \tilde{g}} \, & = & \, \sum_p \frac{N_g}{128\pi} ~ C_p^2 ~ \partial_p (F^{i})^2 \frac{m_{\phi_i}}{M^2_{\rm P}} \nonumber \\
\Gamma_{\phi_i \rightarrow {\tilde G} {\tilde G}} & \sim & \frac{1}{288\pi} ~ \left(|G_{\phi_i}|^2 K_{\phi_i\bar{\phi_i}}^{-1}\right) \frac{m^2_\phi}{m_{3/2}^2}\frac{m_{\phi_i}^3}{M^2_{\rm P}} \,\, . \nonumber \\
\,
\eea
%
where $G =  K +\log |W|^2$ is the K\"ahler function.

Under suitable choices of the K\"ahler potential, the required condition ${\rm Br}_{3/2} \sim 10^{-5}$ may be obtained. Similarly, the decay temperature of the modulus may be obtained in terms of the K\"ahler potential and superpotential. We refer to \cite{Allahverdi:2010rh} for more details.

For a single modulus, the branching ratios to gauge bosons and gauginos are roughly equal, and the branching to the gravitino needs to be suppressed, leading to the condition
\be
\frac{m_{\phi}}{m_{3/2}}|G_{\phi}| K_{\phi \bar{\phi}}^{-1/2} \, \sim \, 10^{-3} \,\,.
\ee

For the KKLT example, the above quantity is $\mathcal{O}(1)$, which leads to overproduction of gravitinos. However, in a more general scenario, one can suppress this ratio to the required levels by a suitable choice of $K_{\phi \bar{\phi}}$ and vacuum expectation value of $\phi$. This does not necessarily affect the existence of other conditions for successful non-thermal Higgsino production, such as comparable anomaly mediated contributions, or a modulus in the correct mass range. Moreover, the scalar masses depend on the holomorphic bisectional curvature of the plane (in tangent space) spanned by the scalars and the supersymmetry breaking modulus \cite{Dutta:2012mw}, and this is not necessarily changed by a shift in the metric $K_{\phi \bar{\phi}}$.
One can therefore expect to have a viable non-thermal scenario with the Higgs mass $m_h \sim 125$ GeV, while suppressing gravitino production from modulus decay. We leave the detailed exploration of these issues for future work.

\section{Conclusions}

In this paper, we have motivated Higgsino dark matter based on a solution to the cosmological moduli/gravitino problem. We revisit the main points of our paper.

Cosmological moduli/gravitino should decay before the onset of BBN, and this requirement sets their minimum mass at roughly $\mathcal{O}(40)$ TeV. Having a gravitino at that mass has implications for the low-energy supersymmetric spectrum, most importantly the gauginos. In different scenarios the gauginos are suppressed with respect to the gravitino and each other in different ways, and while sub-TeV gaugino LSPs may be obtained (for example in pure anomaly mediation), the general ingredients going into a type IIB compactification typically force gauginos above a TeV in this context. Nevertheless, it is possible for the Higgsinos to emerge as sub-TeV dark matter candidates if anomaly and modulus mediated contributions to supersymmetry breaking are both competitive. 

Having obtained sub-TeV Higgsino dark matter, however, we must rely on non-thermal physics to obtain the correct relic density. This in turn places constraints on the decay temperature and mass of the modulus giving rise to the non-thermal history. Moreover, one must ensure that gravitinos arising from modulus decay don't themselves decay into and overproduce dark matter.


We considered the explicit example of mirage mediation model in $D=4,~N=1$ supergravitiy arising from type IIB KKLT compactification, where the modulus decay provides the non-thermal origin of Higgsino-like dark matter. We saw that the annihilation rate of Higgsinos is consistent with the Fermi bounds, and the correct relic density is obtained by non-thermal production. The spin independent scattering cross section, too, is consistent with the latest bounds from direct detection experiments. The large gravitino mass is helpful to yield $m_h$ around $125$ GeV in this model and satisfy the direct detection constraints. Thus, while not motivated explicitly by the Higgs mass, this cosmological approach to the moduli problem and Higgsino dark matter is certainly supported by the Higgs mass.

Within this example, however, the decay of gravitinos overproduces dark matter due to the nature of its coupling to the decaying modulus. We have outlined general requirements for gravitinos to have the necessary coupling to the modulus, while still preserving the other elements in these constructions.




\bibliographystyle{aipproc}   

\bibliography{sample}

\IfFileExists{\jobname.bbl}{}
 {\typeout{}
  \typeout{******************************************}
  \typeout{** Please run "bibtex \jobname" to optain}
  \typeout{** the bibliography and then re-run LaTeX}
  \typeout{** twice to fix the references!}
  \typeout{******************************************}
  \typeout{}
 }

\section{Acknowledgement}

We thank Rouzbeh Allahverdi and Bhaskar Dutta for collaboration and Sheldon Campbell for very useful discussions. We are grateful to the Center for Theoretical Underground Physics and Related Areas (CETUP* 2012) in South Dakota for its hospitality and for partial support during the completion of this work. This  work is partially  supported in part by the DOE grant DE-FG02-95ER40917. 


\end{document}